\documentclass[11pt, preprint]{aastex631}

\usepackage{graphicx}
\usepackage{subfigure}
\usepackage{subfloat}
\usepackage{amssymb}
\usepackage{amsmath}
\usepackage{multirow}
\usepackage{dcolumn}
\usepackage{bm}
\usepackage{wasysym}
\usepackage{cancel}

\shorttitle{Magnetic Effect on Permittivity}
\shortauthors{}

\begin{document}

\title{Magnetic Effect on the Potential Barrier for Nucleosynthesis}

\author{Kiwan Park}
\affil{Department of physics and OMEG Institute, Soongsil University; pkiwan@ssu.ac.kr}
\date{\today}

\begin{abstract}
We demonstrated that a weak magnetic field can increase the permittivity, leading to a reduction in the potential barrier within the Debye sphere consisting of electrons and a nucleus. By solving the Boltzmann equation with the inclusion of the magnetic field, we obtained the magnetized permittivity. The resulting enhanced permittivity field inversely decreases the potential barrier, thereby increasing the reaction rate between two fusing nuclei. We compared this Boltzmann kinetic approach with the Debye potential method. We found that they are qualitatively consistent. Further, we also derived the magnetized Debye potential composed of the conventional term with a new magnetic effect. Both approaches indicate that magnetized plasmas, which have existed since the Big Bang, have ultimately influenced permittivity, potential barrier, and nucleosynthesis.
\end{abstract}


\section{Introduction}
Magnetic fields ($B$) and plasmas are prevalent throughout the Universe. However, despite extensive research, the effect of magnetic fields on the evolution of celestial plasma systems remains partially understood. In particular, the role of magnetized plasma in nucleosynthesis is not well understood. Even a  longstanding debate exists regarding the behavior of unmagnetized plasma (electrons) in fusion ions. Briefly, nucleosynthesis proceeds through a series of processes including the proton-proton (pp) chain, CNO cycle, and triple-alpha reaction. The reaction rate is represented as
\begin{eqnarray}
R\sim \langle \sigma v \rangle &=& \frac{2^{3/2}}{\sqrt{\pi \mu}}\frac{1}{(k_BT)^{3/2}}\int^{\infty}_0 S(E)
\,exp\bigg[-\frac{E}{k_BT}-\frac{Z_1Z_2e^2}{2\epsilon\hbar}\bigg(\frac{\mu}{2 E}\bigg)^{1/2}\bigg]dE \label{Reaction_suppression1}\\
&\sim& \frac{S(E_0)}{T^{2/3}}exp\bigg[  -\bigg(\frac{\mu Z_1^2 Z_2^2}{m_p}\, \frac{7.726\times 10^{10} K}{\epsilon_r T}\bigg)^{1/3} \bigg],
\label{Reaction_suppression2}
\end{eqnarray}
where $\mu$ and $m_p$ are respectively 'reduced mass' and  'proton mass' with $E=3k_BT/2$.  As this formula shows, the nucleosynthesis requires a significant amount of energy to overcome the Coulomb barrier between two fusing ions regardless of quantum tunneling effect. For instance, in the Solar core ($T\sim 10^7 K$) and the early Universe after the Big Bang ($t\sim1-10^2s,\,T\sim 10^{10} K$), the reaction rates for synthesizing deuterium $\mathrm{^2H}$ in the initial step of the proton-proton (pp) chain are suppressed by $1.53\times 10^{-7}$ and $0.21$, respectively. Furthermore, in the subsequent step involving $\mathrm{^3He}$, these rates are further reduced by $1.24\times 10^{-12}$ and $0.064$. So, to explain the ubiquitously existing nuclei, ,i.e., fundamental elements, in the whole Universe, various models have been suggested. And, one of them is the screening effect, attributed to the presence of dense electrons surrounding the ions. The screening effect is believed to lower the Coulomb barrier and enhance the reaction rate (see Debye-Hückel screening \cite{2003phpl.book.....B}). Actually, it is evident that high-density plasmas can reduce the potential barrier from the positive nucleus and elevate the reaction rate. However, such dense plasma state is limited, rather dilute plasmas are more commonly observed.  \\


\cite{1954AuJPh...7..373S} proposed the concept of static electron screening surrounding fusing nuclei, which is essentially equivalent to Debye-Hückel screening. Subsequently, several studies and suggestions were based on this groundbreaking work. \cite{1998PhRvC..57.2756B} solved the Debye potential using the WKB approximation, where the Coulomb wave function naturally emerges from Salpeter's formulation. And, \cite{1998ApJ...504..996G} calculated the partial differential equation for the electron density matrix in the vicinity of two nuclei. Also, Dewitt et al. \cite{1973ApJ...181..439D} and Bruggen et al. \cite{1997ApJ...488..867B} derived the reaction rate based on the free energy between two ions under the assumption of weak screening.\\


Simultaneously, the suitability of Salpeter's static screening effect for dynamic stellar cores was called into question. For example, \cite{1997ASSL..214...43S, 1988ApJ...331..565C, 2021JCAP...11..017H} considered the dynamic effects arising from the disparate velocities of nuclei and electrons. And, \cite{2000ApJ...535..473O} provided a statistical reinterpretation of the Gibbs distribution of particles in plasmas. Furthermore, \cite{1996ApJ...468..433S} investigated the interaction effects of electrons surrounding fusing nuclei. These examples demonstrate that authors have developed their own plasma models based on their respective backgrounds and approaches (\cite{2002A&A...383..291B}, and references therein). Interestingly, however, the influence of the ubiquitous background magnetic field on the permittivity $\epsilon_r=\epsilon/\epsilon_0$ in the penetration factor $P\sim \exp[-g(\epsilon,\,E,\,Z_1,\,Z_2)]$ has not yet been thoroughly explored (refer to Eq.(\ref{Reaction_suppression1})).\\

The magnetic field has existed ubiquitously since the Big Bang. In the very early Universe, various quantum fluctuations, such as QCD or phase transitions followed by plasma fluctuation (Biermann battery effect), induced magnetic fields \citep{1950ZNatA...5...65B, 1994PhRvD..50.2421C, 2012ApJ...759...54T}. These primordial magnetic fields (PMF) are inferred to have been very weak ($10^{-62}-10^{-19}G$) compared to the currently observed mean magnetic field strength ($10^{-5}G$), which implies various dynamo processes. However, the electrons surrounding the nuclei can be magnetized regardless of the strength of the magnetic field. Moreover, a weak magnetic field, which loosely constrains the charged particles (electrons) but still accelerates their motion, can perturb the distribution more efficiently than a strong magnetic field. In contrast, the strong magnetic field has the effect of suppressing the perturbation through the strong constraint.\\

Statistically, the closed structure composed of a nucleus and electrons can be regarded as a canonical ensemble system dominated by Hamiltonian dynamics with generalized coordinates "$q_s$" and momentum "$p_s(=m_sv_s+q_sA)$". Liouville's theorem indicates that the total time (material) derivative of the density or distribution function in phase space is $zero$ as we move along the trajectory dominated by Hamiltonian dynamics. Therefore, some external influences on the system can change the distributions of components, especially light electrons. And it has the effect of modifying the electron density shielding the static electric field from the heavy nuclei. We show that the magnetic effect increases permittivity followed by the drop of the potential barrier between two reacting nuclei.\\

In section 2, we briefly show how to get the permittivity with Boltzmann equation and electromagnetic theory, analytically and numerically. In section 3, we show our numerical results for the magnetized permittivity, potential barrier, and penetration factor. In section 4, we derive the magnetized Debye potential. We used the conventional approach with the additional magnetic effect. In section 5, we summarize our work.

\section{Theoretical Analysis I: Kinetic Approach}
In comparison to the overall distribution $f(\mathbf{r},\, \mathbf{v},\, t)$, the slightly higher-density electrons surrounding the nucleus can be regarded as the perturbed distribution $f_1(\mathbf{r},\, \mathbf{v},\, t)$. Moreover, since the electrons shield the electric field from the nucleus, they effectively act as bound charges $\rho_b=\int f_1(\mathbf{r},\,\mathbf{v},\,t) d\mathbf{r}d\mathbf{v}$ and polarize the system with a dipole moment $\mathbf{P}$: $\mathbf{D}=\epsilon \mathbf{E}=\epsilon_0 \mathbf{E} + \mathbf{P}$. By utilizing the convolution property of Fourier Transformation and taking its divergence, we can separate the longitudinal permittivity $\epsilon_l$ from the electric displacement field $\mathbf{D}$ as follows:
\begin{eqnarray}
k D(k,\,\omega)=k \epsilon(k,\,\omega)E(k,\,\omega)=k\,\epsilon_0E(k,\,\omega)-\rho_b(k,\,\omega)=k\,\epsilon_0E(k,\,\omega)+e\int f_1(\mathbf{r},\,\mathbf{v},\,t)\, d\mathbf{r}d\mathbf{v}.
\label{D_field}
\end{eqnarray}
We apply this relation to the system that is weakly magnetized with $B_0$. The perturbed distribution function $f_1$ is\footnote{Harris dispersion relation is also obtained from this equation \citep{2005mmp..book.....A, 2017ipp..book.....G}. However, Harris mode solves $\epsilon$ for a nontrivial potential in Poisson equation $\nabla^2 \Phi = -\rho/\epsilon_0\rightarrow \Phi(k,\,\omega)=N(k,\,\omega)/D(k,\,\omega)$. With  $D(k,\,\omega)=0(= \epsilon)$, the dispersion relation constraining $k$ and $\omega$ is derived. $\epsilon$ is not permittivity.}
\begin{eqnarray}
\frac{\partial f_{1}}{\partial t}+{\bf v}\cdot \nabla f_{1}-\frac{e}{m_e} {\bf E}\cdot \nabla _V f_{0}-\frac{e}{m_e} {\bf v}\times {\bf B}_0\cdot \nabla _V f_{1}=0.
\label{Boltzmann1}
\end{eqnarray}
Using $v_x=v_{\perp}\cos\,\phi$, $v_y=v_{\perp}\sin\,\phi$, and cyclotron frequency $\omega_{ce}\equiv eB_0/m_e$, we can convert the fourth term, i.e., Lorentz force into $\omega_{ce}\partial f_{1}/\partial \phi$. Then, the Fourier transformed Boltzmann equation is represented as
\begin{eqnarray}
\frac{\partial f_{1}}{\partial  \phi}-i(\alpha+\beta \cos\,\phi)f_{1}+\frac{e}{m_e\omega_{ce}} {\bf E}\cdot \nabla _V  f_{0}=0,\label{Boltzmann3}
\end{eqnarray}
where $\alpha\equiv  (k_{\|} v_{\|}-\omega) / \omega_{ce}$ and $\beta\equiv  k_{\perp} v_{\perp} / \omega_{ce}$. And, then, we get
\begin{eqnarray}
f_{1}=-\frac{e}{m_e\omega_{ce}}\,e^{(\alpha\phi+\beta \sin\,\phi)}\int^{\phi} e^{-(\alpha\phi'+\beta \sin\,\phi')}\,{\bf E}\cdot \nabla_Vf_{0}\,d\phi'.
\label{Boltzmann4}
\end{eqnarray}
Applying $f_1$ to Eq.(\ref{D_field}), we can derive the magnetized permittivity as follows:
\begin{eqnarray}
\epsilon_l=\epsilon_0+\frac{i\epsilon_0}{k^2\omega_{ce}}\omega^2_{pe}\int v_{\perp}dv_{\perp}dv_{\|}d\phi\,\overbrace{e^{i(\alpha\phi+\beta \sin\,\phi)}}^A\int^{\phi} \overbrace{e^{-i(\alpha\phi'+\beta \sin\,\phi')}\,{\bf k}\cdot \nabla_VF_{0}\,d\phi'.}^B
\label{permittivity1}
\end{eqnarray}
Here, plasma frequency $\omega^2_{pe}$ is defined as $n_{e0}q^2_s/\epsilon_0m_e$, and the volume element in cylindrical coordinate is $d^3v=v_{\perp}dv_{\perp}dv_{\|}d\phi$. Also, we use the anisotropic Maxwell distribution $F_{0}=f_{0}/n_{e0}$:
\begin{eqnarray}
F_{0}=\bigg(\frac{1}{2\pi k_BT_{\perp}}\bigg)\bigg(\frac{1}{2\pi k_BT_{\|}}\bigg)^{1/2}e^{-\frac{m_s}{2k_BT_s}(v^2_{\|}+v^2_{\perp})}.
\label{bimaxwellian}
\end{eqnarray}
The exponential term in `$A$' can be represented by Bessel function $ e^{\alpha+\beta \sin\,\phi}=\sum_{m=-\infty}^{\infty}J_m(\beta)e^{i(\alpha+m)\phi}$, and $k\cdot \nabla_V$ is written as $k_{\|}\partial /\partial\,V_{\|}+k_{\bot}\partial /\partial\,V_{\bot}cos\,\phi$ \citep{2003phpl.book.....B, 2005mmp..book.....A}. Combined `$A$' and `$B$' are
\begin{eqnarray}
\sum_{m,\,n}J_m(\beta)J_n(\beta)\bigg[ik_{\|}\frac{\partial F_{0}}{\partial v_{\|}}\frac{e^{i(m-n)\phi}}{\alpha+n}+i\frac{k_{\perp}}{2}\frac{\partial F_{0}}{\partial v_{\perp}}\bigg\{\frac{e^{i(m-n+1)\phi}}{\alpha+n-1}
+\frac{e^{i(m-n-1)\phi}}{\alpha+n+1}\bigg\}\bigg].
\label{Bessel13}
\end{eqnarray}
The index `$n$' is a dummy variable, and $\int^{2\pi}_0 e^{i(m-n)}d\,\phi$ is defined as Dirac delta function $2\pi\delta_{m,\,n}$. Using Bessel recurrence relation  $J_{n+1}(\beta)+J_{n-1}(\beta) = (2\pi / \beta) J_n(\beta)$, we can derive
\begin{eqnarray}
\frac{\epsilon_l}{\epsilon_0}=1+\frac{2\pi\omega^2_{pe}}{k^2}\int_{-\infty}^{\infty}dv_{\|}\int^{\infty}_0v_{\perp}\,dv_{\perp}
\sum_n\bigg[\frac{m_sk_{\|}v_{\|}}{k_BT_e} + \frac{nm_e\omega_{ce}}{k_BT_e}\bigg]\frac{F_{0}J_n^2(\beta)}{k_{\|}v_{\|}-\omega+n\omega_{ce}}.
\label{permittivity4}
\end{eqnarray}
Expanding Bessel function, we can make the result more suitable for the numerical calculation \citep{2005mmp..book.....A}.
\begin{eqnarray}
\frac{\epsilon_l}{\epsilon_0}&=&1+\frac{2\pi\omega^2_{pe}}{k^2}\int_{-\infty}^{\infty}dv_{\|}\int^{\infty}_0v_{\perp}\,dv_{\perp}
\sum_n\bigg[\frac{m_ek_{\|}v_{\|}}{k_BT_e} + \frac{nm_e\omega_{ce}}{k_BT_e}\bigg] \frac{F_{0}}{k_{\|}v_{\|}-\omega+n\omega_{ce}}\bigg(\sum_{e=0}^{\infty}\frac{(-1)^s}{s!(s+n)!}
\bigg(\frac{\beta}{2} \bigg)^{n+2s}\bigg)^2.\nonumber\\
\label{permittivity7}
\end{eqnarray}
Technically, permittivity $\epsilon_l$ in this equation represents the area between the horizontal axis of $v_{\|}$ and integrand. However, the usual residue theorem with singularities cannot be applied because of the divergent $F_{0}$ with $v_{im}\rightarrow \infty$. Instead, we should integrate its principal value and poles directly.
\begin{eqnarray}
\frac{\epsilon_l}{\epsilon_0}&=& 1+\frac{2\pi\omega^2_{pe}}{k^2}P\int_{-\infty}^{\infty}dv_{\|}
\int^{\infty}_0v_{\perp}\,dv_{\perp}\nonumber\\
&&\bigg(\sum_n \frac{m_e}{k_BT_e}\bigg(\frac{k\, v_{\|}\cos\theta
+ n\omega_{ce}}{k\, v_{\|}\cos\theta-\omega+n\omega_{ce}}\bigg)F_{s0}\bigg[\sum_{s=0}^{\infty}\frac{(-1)^s}{s!(s+n)!}
\bigg(\frac{m_ekv_{\perp}\sin\theta}{2eB_0}\bigg)^{n+2s}\bigg]^2\bigg)\nonumber\\
&& + i\pi \frac{k}{|k|}\frac{2\pi\omega^2_{pe}}{k^2}
\sum_n\frac{m_e}{k_BT_e}\big(k\, v_{\|}\cos\theta+nm_e\omega_{ce}\big)\bigg[\sum_{s=0}^{\infty}\frac{(-1)^s}{s!(s+n)!}
\bigg(\frac{m_ekv_{\perp}\sin\theta}{2eB_0} \bigg)^{n+2s}\bigg]^2F_{0}.
\label{final_epsilon_numerical}
\end{eqnarray}
We applied trapezoidal rule to calculate Eq.(\ref{final_epsilon_numerical}) but did not consider the imaginary part in this paper \citep{2012phpl.book.....M}. The range of wavenumber $k$ is from $1$ to $3000$, $v_{min}=-10^8$ to $v_{max}=10^8$, and mesh size is $\Delta v=0.5$. We inferred the electron density $n_{e0}=8.18363\times 10^{13}m^{-3}$ and temperature $T_e=2.38\times 10^6\,K$ near the Solar tachocline regime with arbitrary frequency $\omega=10^4 Hz$ smaller than the plasma frequency $\omega_{pe}=5.1\times 10^8 Hz$. The numerical range within 30\% of light velocity $c$ is large enough to normalize the distribution function with the mesh. Also, the wavenumber is sufficient for the discrete Fourier transform. We divided the integral range of $v_{\|}$ into ($v_{min},$ $v_{res,\,n}-\delta$) and ($v_{res,\,n}+\delta,\, v_{max}$) skipping the singular point $v_{res,\,n}$  with $\delta=5\times 10^{-7}m$. We expanded Bessel function up to $v^{18}_{\perp}$ for the case that the velocity is almost parallel to the $B$ field, i.e., $\beta\sim v_{\bot}\sim0$. And, the result was already saturated in the order of $v^{10}_{\perp}$.\\


\section{Numerical result}
\begin{figure*}
    {
   \subfigure[$\epsilon(k)$ (Eq.\ref{final_epsilon_numerical})]{
     \includegraphics[width=10 cm]{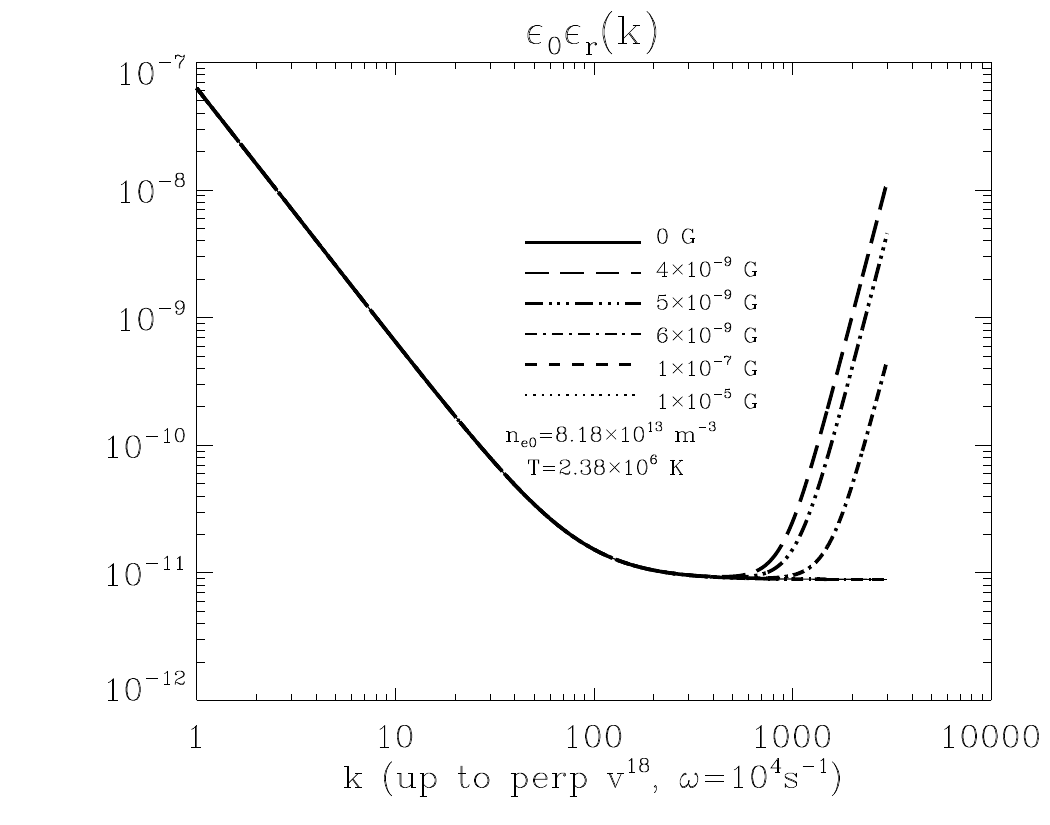}
     \label{f1}
    }\hspace{-15 mm}
   \subfigure[ $\epsilon(r/\lambda_D)$ ]{
     \includegraphics[width=10 cm]{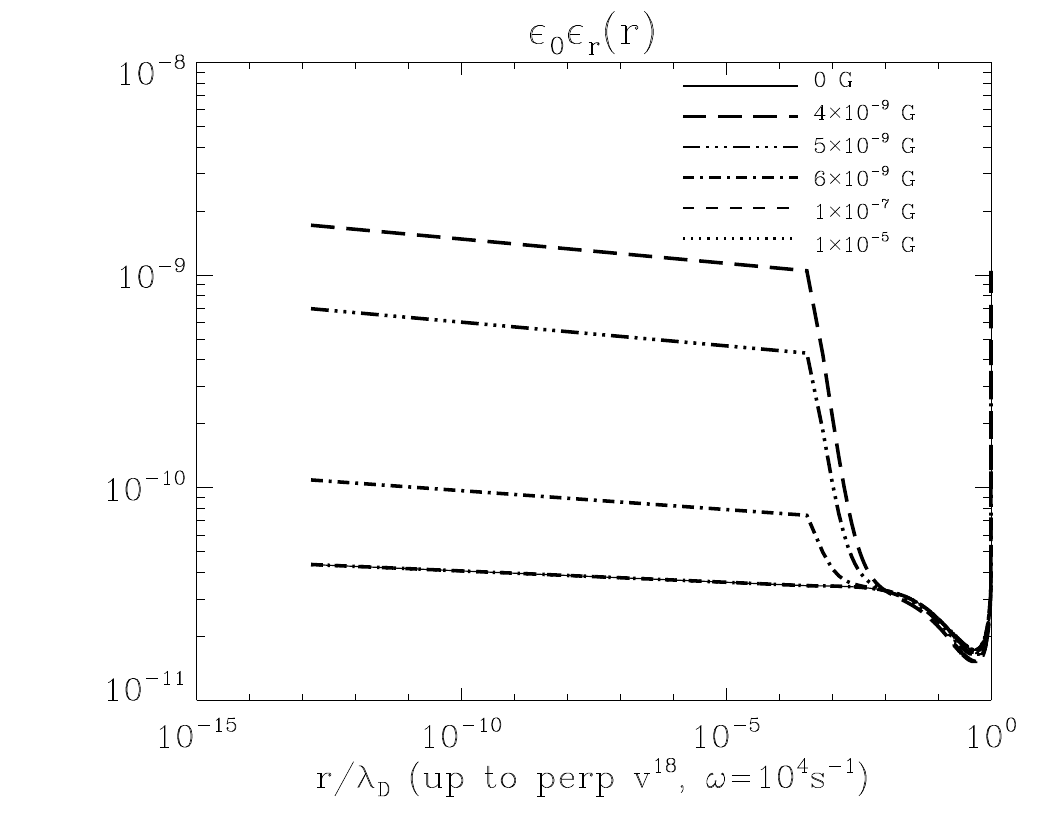}
     \label{f2}
   }\hspace{-15 mm}
   \subfigure[$\frac{e}{4\pi \epsilon r}$]{
     \includegraphics[width=10 cm]{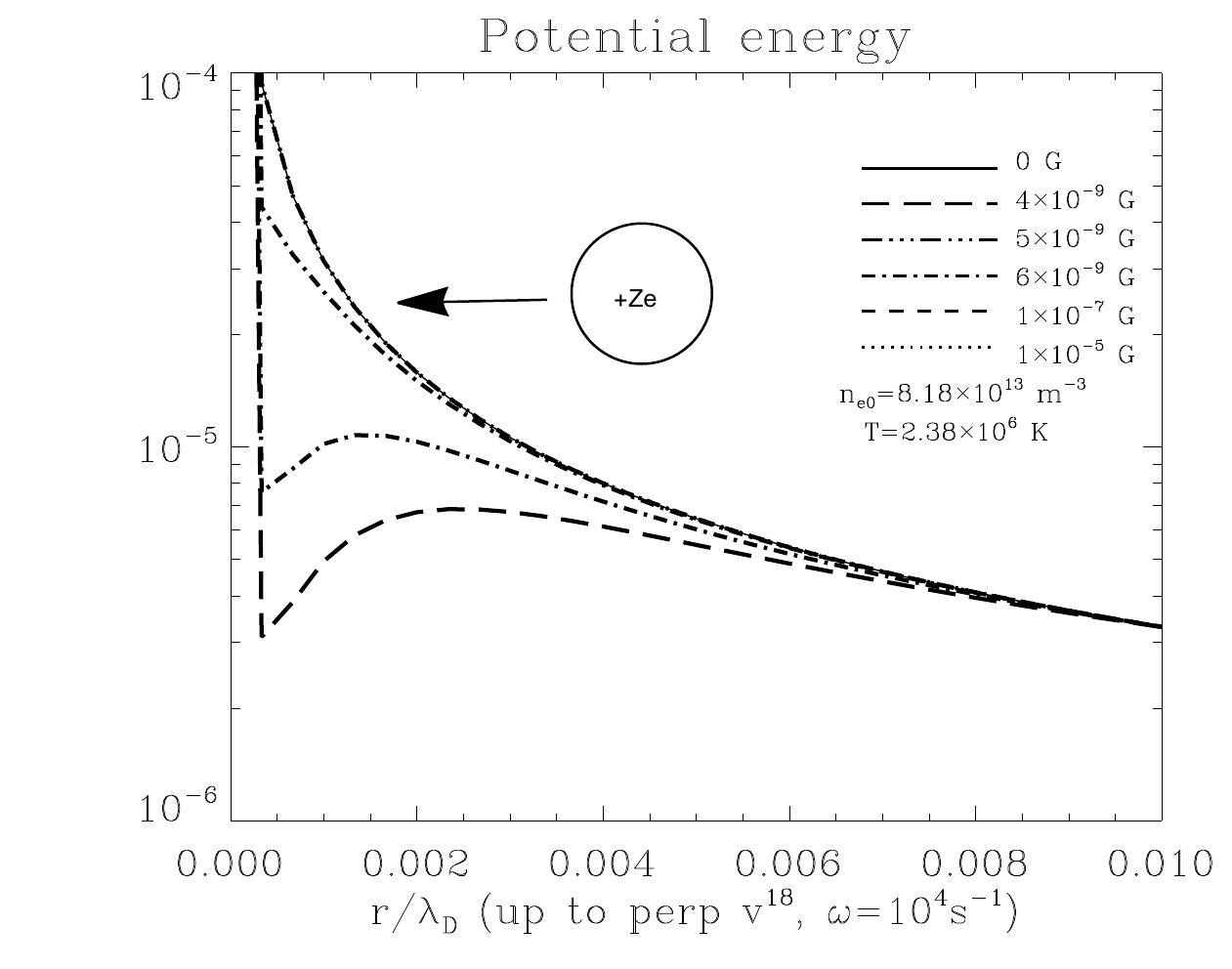}
     \label{f3}
   }\hspace{-15 mm}
   \subfigure[$exp\bigg(-\frac{Z_1Z_2e^2}{2\epsilon \hbar}\sqrt{\frac{\mu}{2E}}\bigg)$, $Z_1=Z_2=1$ ]{
     \includegraphics[width=10 cm]{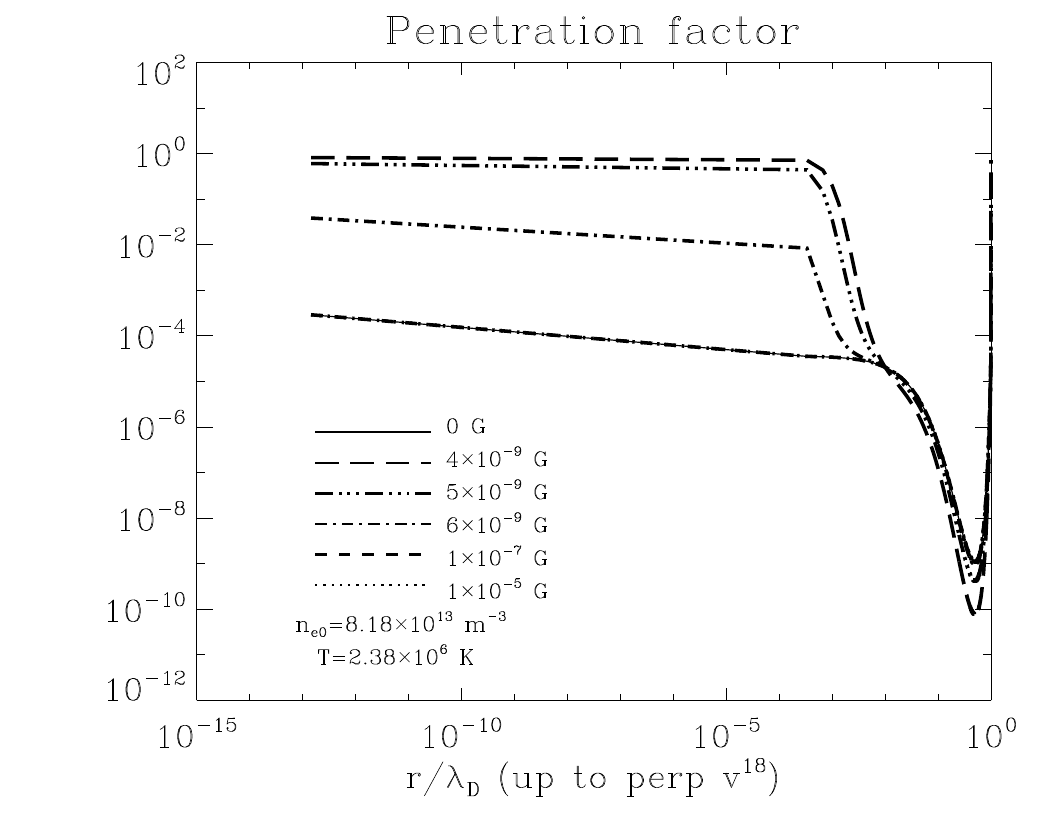}
     \label{f4}
   }
}
\caption{The $B$ field of $1\times 10^{-5}G$(dotted line) in fact makes the same result as that of $B=0G$ (solid line).}
\end{figure*}

Fig.\ref{f1} illustrates the Fourier-transformed evolving permittivity ($\epsilon_0\epsilon_r$, $\epsilon_0=8.85\times 10^{-12} F/m$) influenced by the magnetic field. We applied various magnetic fields ($0-1\times 10^{-5}{G}$) to the system with wavenumbers $k$ ranging from $1$ to $3000$. The permittivity remains degenerate up to a certain critical wavenumber $k_{{crit}}$, regardless of the magnetic field strength. However, it becomes separated for $k > k_{crit}$ and is amplified by the weak magnetic field. The degree of separation is inversely proportional to the strength of the magnetic field. For magnetic fields stronger than $1\times 10^{-7}G$, the permittivity becomes essentially the same as the unmagnetized case. This can be attributed to the $B$ term in the denominator of Eq.(\ref{final_epsilon_numerical}). Furthermore, Eq.(\ref{Boltzmann1}) demonstrates the growth of $f_1$ as the Lorentz force decreases, which is consistent with Liouville's theorem. Fig.\ref{f2} shows $\epsilon_r(r/\lambda_D)$ in real space. Here, `$r$' represents the distance from the nucleus, $\lambda_D$ is Debye length $\sqrt{\epsilon_0k_BT/e^2n_e}\sim 1.17\times 10^{-2}cm$. We performed an inverse Fourier transform of $\epsilon(k)$ using
\begin{eqnarray}
\epsilon(r_n)=\frac{1}{N}\sum_{k=0}^{N-1}\epsilon(k)exp\bigg(i\frac{2\pi k n}{N}\bigg),
\label{IDFT1}
\end{eqnarray}
where $r_n=nL/N$. Near the nucleus, $\epsilon_r$ is split into the various levels according to the applied $B$ field, inversely proportional to the magnetic field. However, above the critical $B_{crit}$ field, permittivity is not split but converges to the nonmagnetized case. And, at $n\sim N-1$, the oscillation by $\mathrm{cos} (2\pi k n/N)$ is almost negligible, which appears as the sudden increase of $\epsilon(r)$ at $r \sim \lambda_D$. Fig.\ref{f3} includes the evolution of potential energy $\phi=Q/4\pi \epsilon r$ for a hydrogen nucleus with permittivity $\epsilon(r)$. Since permittivity constitutes the denominator, potential energy evolves in response to the $B$ field. The plot illustrates that the weak magnetic field decreases the potential barrier. Fig.\ref{f4} shows the evolution of penetration factor $P(E)$. The result clearly show that the weak $B$ field, which reduces the potential barrier, enhances the probability of penetration and reaction. The actual potential barrier is of course much more complex. In principle. it should be calculated with the interaction energy among the screening charges around two interacting nuclei and environmental lighter nuclei. However, we do not their effects at the moment. We focus on the weak magnetic effect on the perturbed distribution $f_1$, bound charges, and the enhanced nucleus reaction. This may be a more common mechanism in the whole Universe history.\\

\begin{figure*}
{
   \subfigure[]{
     \includegraphics[width=10 cm]{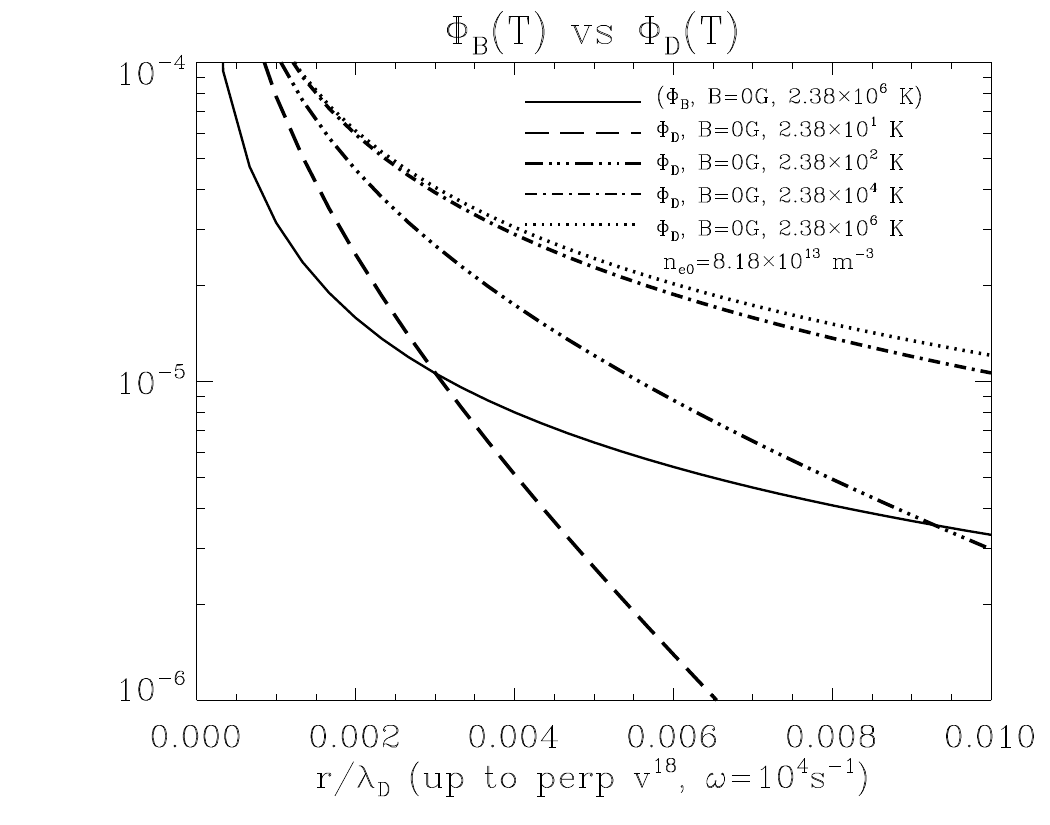}
     \label{f5}
}\hspace{-15 mm}
   \subfigure[]{
     \includegraphics[width=10 cm]{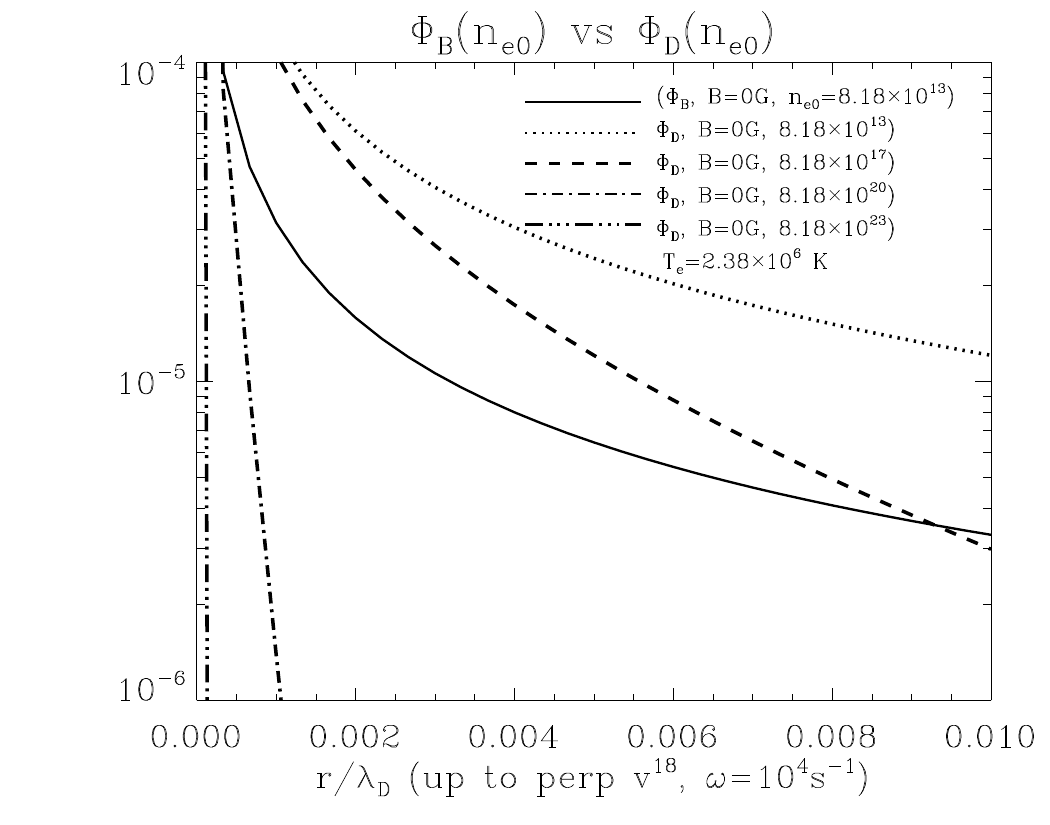}
     \label{f6}
   }
}
\caption{The solid lines indicate potential energy from Eq.(\ref{Boltzmann1}). Other lines indicate Debye potential with various conditions.}
\end{figure*}

In Fig.~2, we compared the kinetic approach with the conventional Debye screened potential. In the absence of the magnetic field ($B_0=0$), the perturbed electron density around the nucleus can be simply represented as
\begin{eqnarray}
f_{1}=\frac{e}{i\,m_e}\frac{{\bf E}\cdot \nabla_Vf_{0}}{({\bf k}\cdot{\bf v}-\omega)}.
\label{Boltzmann_no_B}
\end{eqnarray}
This equation can be calculated using the same method as in Eq.(\ref{Boltzmann1})-(\ref{final_epsilon_numerical}) and compared to Debye potential:
\begin{eqnarray}
\phi=\frac{Q}{4\pi \epsilon_0r}\exp\bigg[-\frac{\sqrt{2}r}{\lambda_D}\bigg],\,\, \epsilon\rightarrow \epsilon_0\exp\bigg[r \sqrt{\frac{2n_e e^2}{\epsilon_0k_BT_e}}\bigg].
\label{debye_potential1}
\end{eqnarray}
Fig.\ref{f5} illustrates that potential energy increases with the increasing temperature. However, potential energy in Fig.\ref{f6} decreases as the electron density increases. The nonmagnetized potential energy from Eq.(\ref{Boltzmann_no_B}) is qualitatively consistent with the Debye potential. However, if there is a current density ${\bf J}=Nq_e{\bf V}$ present, the dependence of potential energy on the electron density becomes opposite to our case and the Debye screening potential energy \cite{2000PhPl....7.3476B, 2013leel.book.....D}. The kinetic model and Debye approach explicitly and implicitly assume the presence of bound charge rather than the current density $\bf J$.\\

\section{Theoretical Analysis II: Magnetized Debye potential}
We can also consider the magnetic effect on the conventional Debye potential. The momentum equation with the $B$ field and collision frequency $\nu_m$ is represented as
\begin{eqnarray}
\frac{d\mathbf{V}_s}{dt}&=&q_s({\bf E}+{\bf V}_s\times {\bf B})-\frac{k_BT}{n_s}\nabla n_s-\nu_m{\bf V}_s\\
\rightarrow -i\omega V_s&\sim &-q_s\nabla\Phi+q_sV_s B-\frac{k_BT}{n_s}\nabla n_s-\nu_mV_s.\,\,(s=i,\,e)
\label{derivation_of_new_Debye_potential_1}
\end{eqnarray}
We integrate the equation from $\infty$ to $r$. Then,
\begin{eqnarray}
&&n_i(r)=n_0\,\exp\big[ -\frac{e\Phi(r)}{k_BT} + \frac{1}{3eB}(i\omega-\nu_m+eB)\big],\\
&&n_e(r)=n_0\,\exp\big[ +\frac{e\Phi(r)}{k_BT} -\frac{1}{3eB}(i\omega-\nu_m-eB)\big].
\label{derivation_of_new_Debye_potential_3}
\end{eqnarray}
Here, we used the dimensional analysis and mean value theorem assuming the quasi-continuous velocity distribution:
\begin{eqnarray}
&&n_e(\infty)=n_i(\infty)\equiv n_0, \int^r_{\infty}\nabla \Phi\,dr=\Phi(r), \\
&&\int^r_{\infty}V_idr\sim U_i(r) - {U_i(\infty)} \rightarrow \overline{U}_i\overline{r}_i,\,\,
\int^r_{\infty}V_e dr\sim {U_e(r)} - U_e(\infty)\rightarrow - \overline{U}_e\overline{r}_e
\label{derivation_of_new_Debye_potential_assumptions_0}
\end{eqnarray}
Additionally, we used $k_BT=3m_s\overline{U}^2_s$ and $\overline{r}_s=m_s\overline{U}_s/q_sB$ based on the balance between Lorentz force and centrifugal force. It should be noted that $\overline{U}_s$ with $m_s$ for $s=i,\,e$ was replaced by the system temperature $T$. Subsequently, by applying $-\epsilon_0 \nabla^2 \Phi=e(n_i-n_e)$, the potential energy can be represented as follows:
\begin{eqnarray}
\frac{1}{r^2}\frac{\partial}{\partial \,r}\bigg(r^2 \frac{\partial\,\Phi}{\partial\,r}\bigg)-\frac{2}{\lambda^2_D}\Phi+\frac{2n_0}{3\epsilon_0B}\,(i\omega-\nu_m)=0.
\label{derivation_of_Debye_potential_4}
\end{eqnarray}
With a trial function $\Phi=QF(r)/4\pi \epsilon_0 r$, we have
\begin{eqnarray}
\Phi(r)=c_1\frac{Q}{4\pi \epsilon_0 r}e^{-\frac{\sqrt{2}r}{\lambda_D}}+c_2\frac{Q}{4\pi \epsilon_0 r}e^{\frac{\sqrt{2}r}{\lambda_D}}+\frac{n_0\lambda^2_D}{3\epsilon_0 BQ}(i\omega-\nu_m).
\label{derivation_of_Debye_potential_6}
\end{eqnarray}
Since $\nu_m$ is caused by the combined effect of electric field, magnetic field, and thermal pressure, the collision frequency may be limited to the internal range of $\lambda_D$. So, in order to satisfy $\Phi(\infty)=0$ for $r\gg \lambda_D$, $c_2$ can be
\begin{eqnarray}
c_2=-\frac{4\pi r}{Q}e^{-\sqrt{2}r/\lambda_D}\frac{i\omega\,n_0\lambda^2_D}{3BQ}.
\label{c_2_derivation_of_Debye_potential_6}
\end{eqnarray}
If we set $c_1=1$ for consistency, the modified potential is
\begin{eqnarray}
\Phi(r)=\frac{Q}{4\pi \epsilon_0 r}e^{-\frac{\sqrt{2}r}{\lambda_D}}-\frac{n_0\nu_m\lambda^2_D}{3\epsilon_0 BQ}.
\label{derivation_of_Debye_potential_7}
\end{eqnarray}
This result demonstrates that the potential is proportional to the magnetic field $B$ and is consistent with the kinetic model. The magnetic effect becomes evident with the balance between the centrifugal force and Lorentz force. Detailed information on $\nu_m$ is required for more exact investigation. We will not delve further into this topic at present. Nonetheless, note that the magnetic field plays as if it were an additional charged particle in the modified Debye potential.\\

\section{Summary}

In our study, we solved the weakly magnetized Boltzmann equation by considering the system as an isolated canonical ensemble composed of the nucleus and bound charges. We demonstrated that the permittivity is inversely proportional to the magnetic field, indicating that the potential barrier between two fusing nuclei evolves proportionally with the magnetic field. This result is related with Liouville theorem, which states that the net change of density or distribution function in phase space is zero as we move along the trajectory dominated by Hamiltonian dynamics. The weak magnetic field reduces the acceleration effect in the Boltzmann equation, resulting in a decreased constraint on electrons by the magnetic field. This leads to an enhanced fluctuating electron distribution $f_1(\mathbf{r},\,\mathbf{v},\,t)$ in configuration space to compensate for the loss. The equation $\nabla \cdot (\epsilon {\bf E})=\epsilon_0 \nabla \cdot {\bf E}+e\int f_1d\mathbf{r}d\mathbf{p}$ explains how the growth of $f_1$ contributes to the increasing permittivity, which in turn leads to a decrease in the potential barrier. In contrast, for magnetic fields beyond a critical threshold, the electrons are strongly constrained, causing the system to behave similarly to a non-magnetized system. It is worth noting that in addition to the magnetic field, the permittivity is also influenced by factors such as electron density, temperature, and current density \citep{1968SvPhU..10..509V, 2000PhPl....7.3476B, 2017inel.book.....G}. The modified permittivity resulting from these effects ultimately affects the reaction rate in nucleosynthesis in the Universe.\\

The authors acknowledge the support from National Research Foundation of Korea:NRF-2021R1I1A1A01057517, NRF-2020R1A2C3006177, NRF-2021R1A6A1A03043957, and NRF-2020R1F1A1072570.

\bibliographystyle{aasjournal}
\bibliography{bibfile0329}

\end{document}